\documentclass[useAMS,usenatbib]{mnras}
\usepackage{times}
\usepackage{latexsym, amssymb, amscd, psfrag}
\usepackage[dvipdfmx]{graphicx} 
\usepackage{bmpsize}
\usepackage{epsfig,color}
\usepackage{subfig}
\usepackage{morefloats,rotating,float}
\usepackage{gensymb}
\usepackage{multirow,array}
\usepackage{appendix}

\title[GAMA: Red spiral galaxies]{Galaxy And Mass Assembly (GAMA): Properties and evolution of red spiral galaxies}
\author[Mahajan et al.]
{\parbox{\textwidth}{Smriti Mahajan$^{1}$\thanks{E-mail: \texttt{mahajan.smriti@gmail.com}}, Kriti Kamal Gupta$^{1}$, Rahul Rana$^{1}$, M. J. I. Brown$^{2}$, S. Phillipps$^{3}$, 
Joss Bland-Hawthorn$^{4}$, M. N. Bremer$^{5}$, S. Brough$^{6}$, B.W. Holwerda$^{7}$, A. M. Hopkins$^{8}$, J. Loveday$^{9}$, Kevin Pimbblet$^{10}$, Lingyu Wang$^{11,12}$
 }\vspace{0.4cm} \\
 \parbox{\textwidth}{$^{1}$Indian Institute of Science Education and Research Mohali- IISERM, Knowledge City, Manauli, 140306, Punjab, India \\
 $^{2}$School of Physics and Astronomy, Monash University, Clayton, VIC 3800, Australia \\
 $^{3}$Astrophysics Group, School of Physics, University of Bristol, Tyndall Avenue, Bristol BS8 1TL, UK \\
 $^{4}$Sydney Institute for Astronomy, School of Physics A28, University of Sydney, NSW 2006, Australia \\
 $^{5}$Astrophysics Group, School of Physics, University of Bristol, Tyndall Avenue, Bristol BS8 1TL, UK \\
 $^{6}$School of Physics, University of New South Wales, NSW 2052, Australia \\
 $^{7}$Department of Physics and Astronomy, 102 Natural Science Building, University of Louisville, Louisville KY 40292, USA \\
 $^{8}$Australian Astronomical Optics, Macquarie University, 105 Delhi Rd, North Ryde, NSW 2113, Australia \\
 $^{9}$Astronomy Centre, University of Sussex, Falmer, Brighton BN1 9QH, UK \\
 $^{10}$E.A.Milne Centre for Astrophysics, University of Hull, Cottingham Road, Kingston-upon-Hull, HU6 7RX, UK \\
 $^{11}$SRON Netherlands Institute for Space Research, Landleven 12, 9747 AD, Groningen, The Netherlands \\
 $^{12}$Kapteyn Astronomical Institute, University of Groningen, Postbus 800, 9700 AV, Groningen, The Netherlands 
}  } 

\def\g{{GAMA }}

\def\gr{{$(g-r)^0$}}
\def\smass{{$M^*/M_\odot$}}
\def\smassa{{$M^*$}}
\def\arcsec{{^{\prime\prime}}}
\def\dmass{{$M_{dust}$}}
\def\hmass{{$M_{HI}$}}
\def\m{{$\mu$}}
\def\w{{\it WISE}}

\definecolor{grey}{rgb}{0.5,0.6,0.7}

\definecolor{amber}{rgb}{1.0,0.49,0.0}

\begin{document}

\date{}
\pubyear{}


\pagerange{\pageref{firstpage}--\pageref{lastpage}} 
\maketitle

\label{firstpage}
\begin{abstract}

We use multi-wavelength data from the Galaxy and Mass Assembly (GAMA) survey to explore the cause of red optical colours in nearby 
 ($0.002<z<0.06$) spiral galaxies. We show that the colours of red spiral galaxies are a direct consequence of some environment-related mechanism(s) which has removed dust and gas, leading to a 
 lower star formation rate. We conclude that this process acts on long timescales (several Gyr) due to a lack of morphological transformation associated with the transition in optical colour.
 The sSFR and dust-to-stellar mass ratio of red spiral galaxies is found to be statistically lower than blue spiral galaxies. On the other hand, red spirals are on average $0.9$ dex more massive, 
 and reside in environments 2.6 times denser than their blue counterparts. We find no evidence of excessive nuclear activity, or higher inclination angles to support these as the major causes for the 
 red optical colours seen in $\gtrsim$ 47\% of all spirals in our sample. Furthermore, for a small subsample of our spiral galaxies which are detected in HI, we find that the SFR of gas-rich red spiral galaxies 
 is lower by $\sim 1$ dex than their blue counterparts.
\end{abstract}

\begin{keywords}
  galaxies: evolution; galaxies: fundamental parameters; galaxies: structure; galaxies: star formation; galaxies: stellar content
\end{keywords}

 \section{Introduction}
 \label{intro}
  
 Conventional wisdom suggests that most spiral galaxies are vigorously forming stars leading to blue optical colours, and are younger relative to their elliptical 
 counterparts. Spiral galaxies also preferentially inhabit less dense environments \citep[e.g.][]{dressler80} relative to passively evolving elliptical galaxies. 
 Large datasets from all-sky surveys have however challenged this view by establishing the existence of red spiral galaxies. Red spiral galaxies may result from attenuation due to dust 
 \citep[e.g.][]{valentijn90,driver07,koyama11}, high metallicity \citep[e.g.][]{mahajan09}, low star formation rate (SFR) \citep[e.g.][]{moran06,masters10b,cortese12b,tojeiro13} or no 
 detectable star formation \citep{fraser16}. 
 
 Red optical colours of a significant fraction of spiral galaxies in the nearby Universe is a debatable issue in the literature.  \citet{mahajan09} showed that for a 
 sample of $\sim 6,000$ galaxies found in low redshift ($0.02<z\leq 0.12$) clusters, only $\sim 50\%$ of
 the red star-forming galaxies are dust-reddened. \citet{mahajan09} found statistical evidence favouring high metallicity as the cause of optical
 red colours in the remaining half of the star-forming galaxies in and around clusters.  On the other hand, a recent study  \citep{bremer18} of nearby galaxies, 
 found in the green valley of the optical colour-magnitude space showed that the change in colours from blue to red is primarily due to the change in the colour of the disc.
 \citet{bremer18} also showed that most of the green valley galaxies have significant bulges, suggesting that a strong bulge is present prior to decline in the star formation of galaxies. 
 
 Several other studies made use of multi-wavelength data to show that the star 
 formation activity of optically-red spiral galaxies is not different from their blue counterparts \citep[e.g.][]{cortese12a,bonne15}. 
 But \citet{tojeiro13} found that star formation in the red spirals is reduced by a factor of three relative to their blue counterparts in the last $\sim 500$ Myr.
 These authors also found that the star formation history of red and blue spirals is similar at earlier times, and that red spirals are still forming stars $\sim 17$ times faster
 than red ellipticals over the same period of time. These observations are also in broad agreement with the recent findings from submillimeter surveys which have
 discovered populations of optically-red, massive star-forming galaxies, likely to be the recent progenitors of red, passive galaxies at $z\sim0$ \citep[e.g.][]{eales18a,eales18b}.

 Large inclination angles cause large optical depth, which may result in the unexpected red colours of spiral galaxies \citep[e.g.][]{valentijn90,driver07}. Furthermore, 
 the total extinction in spirals in the $r$-band is found to increase from face-on to edge-on spirals by 0.5 mag \citep{masters10a}. \citet[]{koyama11} found many dusty red H$\alpha$ emitters around 
 Abell 851 ($z=0.4$) associated with groups of galaxies \citep[also see][]{santos13}. This discovery supports the scenario where pre-processing in groups involves 
 dusty star formation activity, which eventually truncates star formation in infalling galaxies \citep{koyama11,mahajan12,mahajan13}. But other authors \citep[e.g.][]{masters10b} 
 found no correlation between optical colours and the environment of face-on disks, which led them to conclude that environment alone can not transform the optical colour of a galaxy. 
     
 It has also been suggested that the red optical colours of some spirals could result from the presence of nebular emission in these galaxies \citep{masters10b,kaviraj15}. 
 In an independent analysis of HI-detected, passive galaxies, \citet{parkash19} find that the integral field unit spectra of 20 out of 28 galaxies in their sample have extended low-ionization
 emission-line regions (LIERs) and 1 has low-ionization nuclear emission-line regions (LINERs). \citet{parkash19} therefore concluded that 75\% of HI galaxies with little or no star formation are LIERs
 or LINERs. On the contrary, in a pilot study of six passive spiral galaxies ($z<0.035$) using multi-band photometric and integral field spectroscopic data, \citet{fraser16} found that 
 none of the galaxies in their sample showed signs of substantial nebular emission.

 In this work we make use of data products obtained from a multi-wavelength dataset covering ultraviolet (UV) to 21 cm radio continuum observations available for a variety of galaxies in the nearby 
 Universe, to address the cause of red optical colour of some spiral galaxies. Our goal is to (i) compare an unbiased sample of red spiral galaxies with their blue counterparts 
 using various physical properties, in order to get an insight into the cause of red optical colours, and (ii) check if the optical colour of these spirals is reversible, i.e. 
 if the red spirals have enough fuel to form new stars, which can push them back into the blue cloud in the colour-magnitude plane. 
 We present the dataset and the derived properties of galaxies analysed in this work in the following section. In Sec. 3 we compare various properties of the red and blue
 spiral galaxies to establish how the two populations differ beyond optical colours. We then incorporate the 21 cm HI continuum data to get further insight
 into the properties of spiral galaxies, and explore the dust and gas content of red spiral galaxies in Sec. 4. Finally, we discuss our observations in the light of 
 the existing literature in Sec. 5 and present our final conclusions in Sec. 6. Throughout this work we use concordance $\Lambda$ cold dark matter cosmological model
 with $H_0 = 70$ km s$^{-1}$ Mpc$^{-1}$, $\Omega_\Lambda = 0.7$ and $\Omega_m = 0.3$ to calculate distances and magnitudes. 
    
 \section{data}
 \label{data}

 In this paper we utilise the $r$-band selected sample of galaxies from the Galaxy and Mass Assembly (GAMA) survey 
 \citep{driver11,hopkins13,liske15}.
 GAMA is a multi-wavelength campaign based on the SDSS (Data Release 7) photometry, and has obtained spectra for $\sim 3$ million galaxies
 with the AAOmega spectrograph on the Australian Astronomical Telescope (AAT). This work is based on the data from the three equatorial regions
 (G09, G12 and G15) covering $180$ square degrees on the sky. GAMA is $>98\%$ complete to $r\leq19.8$ mag \citep{driver11}. 

 The parent sample for this work was selected from the LocalFlowCorrection data management unit (DMU henceforth) version 14 \citep{baldry12}. 
 We selected all galaxies with $NQ>2$ \citep{baldry10} in the redshift range $0.002-0.06$. The former criterion ensures a high quality redshift for the
 selected galaxies, while the latter is chosen to exclude Galactic stars and obtain Hubble type classification following \citet{kelvin14} from the 
 VisualMorphology DMU. These criteria result in a master sample of 7,984 galaxies.
 
 \subsection{Spectroscopic and photometric data}
 
 The matched aperture photometry for all galaxies in 21 wavebands is obtained from the LambdarPhotometry DMU version 1
 \citep{wright16}. The Lambda Adaptive Multi-Band Deblending Algorithm in {\sc R} ({\sc LAMBDAR}) calculates the matched aperture 
 photometry across images that are neither matched in pixel-scale nor point spread function, using prior aperture definitions 
 derived from high-resolution optical imaging. Specifically, we use the LambdarInputCatUVOptNIR v01, 
 LambdarSDSSgv01 and LambdarSDSSrv01 catalogues from the 
 LambdarPhotometry DMU. The magnitudes are then $k-$corrected to $z=0$ using the prescription of \citet{chilingarian12}. The median $k$-correction for our sample is 
 $\sim 0.03$ mag in the $g$ and $r$ bands. We have also made use of the \w~photometry from WISEcatv02 \citep{cluver14} for our sample.
 
 \begin{figure*}
 \centering{
     {\epsfig{file=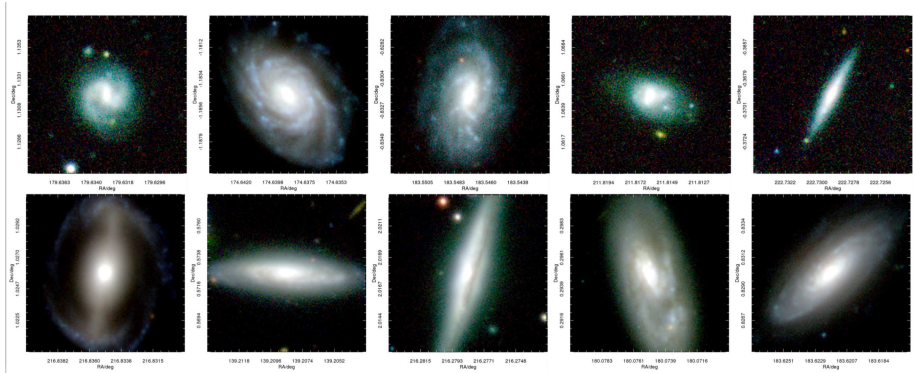,width=12cm}}
  }
 \caption{This figure shows composite images from the Visible and Infrared Survey Telescope for Astronomy (VISTA) Kilo Degree survey (KIDS) in the {\it g,r,i} filters of randomly 
 chosen blue {\it (top)} and red {\it (bottom)} spiral galaxies from our sample. Each image is 20 square arcsecond. Despite very similar spiral morphologies, the two samples show different 
 colours.  } 
 \label{images}
 \end{figure*}
 
 \subsection{The spiral galaxies}
 \label{spirals}
 
 Our sample of spiral galaxies is based on the classification provided in the VisualMorphology DMU (version 3). Specifically, we adopted
 the Hubble type classification and selected all galaxies which are classified as Sa, Sb or Sc type spirals (Table~\ref{spiral}). Two authors (KKG and RR) further
 visualized the 5-colour SDSS images of all galaxies in our sample classified as Sd-Irr to find 55 Sd spirals and 3,276 irregular galaxies.
 Another 208 galaxies in this class remain unclassifiable into either of these two categories. 
 Since the number of unclassified Sd-Irr galaxies is small, we decided to incorporate them in the following analysis along with the 55 Sd spirals. We also repeated our analysis 
 by excluding the unclassifiable galaxies, and confirm that all the results presented here are statistically robust against exclusion of the 
 unclassifiable galaxies in the final sample.
 
 Therefore, our final sample of spirals comprises 2,512 galaxies (including a small fraction of irregular galaxies and lenticulars),
 which is henceforth referred to as `spirals' for convenience and is used throughout this paper unless specified otherwise.
 
  \begin{table}
 \caption{Morphological classification of our sample of spiral galaxies.}
 \begin{center}
 \begin{tabular}{ l|l|r }     
 \hline
      Class  & Description & Number of galaxies  \\ \hline
     S0-Sa  & Lenticulars and spirals & 746 \\
     SB0-SBa & Lenticulars and spirals & 80 \\
     Sab-Scd & Spirals & 1,232 \\
     SBab-SBcd & Spirals & 191 \\
     Sd-Irr & Spiral and irregular galaxies & 3,539 \\
 \hline
 \end{tabular}
 \end{center}
 \label{spiral}  
 \end{table}

 \begin{figure}
 \centering{{\rotatebox{270}{\epsfig{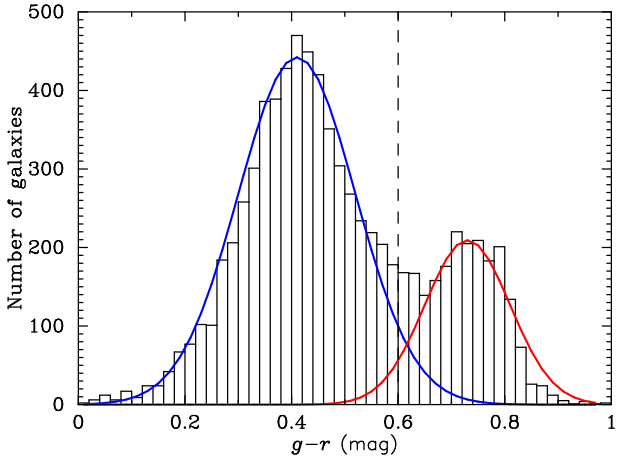}}}}
 \caption{The distribution of $(g-r)$ colour for all GAMA galaxies ($0.002<z<0.06$). Two Gaussians are fitted to this distribution in order to classify 
 galaxies on the basis of colour. The critical value obtained is $(g-r)=0.6$, such that all galaxies with $(g-r)>0.6$ mag are considered as red in this 
 work.  
 } \label{gr}
 \end{figure}

 Figure~\ref{gr} shows the bimodal $g-r$ colour distribution of all GAMA galaxies ($0.002<z<0.06$). In order to 
 divide our sample into red and blue, we fit the colour distribution with two gaussians with mean (standard deviation, $\sigma$) at 0.41 (0.10) and
 0.73 (0.08). Based on this exercise we adopt the colour value $(g-r)=0.6$ mag, which is $1.7\sigma$ from the mean of both the fitted gaussians, as the boundary between the red and 
 blue spiral galaxies. This criteria results in 2,203 red
 galaxies ($g-r>0.6$) in the master sample, of which 1,049 are spirals. Some examples of spiral galaxies in the red and blue sub-samples are shown in Fig.~\ref{images}. 
 This is in broad agreement with the study of \citet{bonne15}, who used the Two Micron All Sky Survey Extended Source Catalog to report the
 fraction of red spirals to be 20\%--50\% of all spirals with $-25 \leq M_K < -20$, and in excess of 50\% for galaxies brighter than $M_K < -25$ mag.
 
 In the following, we compare the physical properties of red spirals with their blue counterparts, and the ensemble of all GAMA galaxies in our 
 chosen redshift range. 

 \subsection{Physical properties}
 
 Physical parameters for all GAMA galaxies in the three equatorial regions have been obtained by fitting the 21-band photometric data with the
 stellar energy distribution code Multi-wavelength Analysis of Galaxy Physical Properties \citep[{\sc magphys};][]{dacunha08}. The {\sc magphys}
 output includes star formation rate (SFR), specific star formation rate (SFR/$M^*$; sSFR), stellar mass ($M^*$), dust mass ($M_{dust}$),
 $r$-band light weighted age and metallicity amongst others. We compile these properties for our sample from the Magphys DMU (version 6). 
 
 The SFR obtained by {\sc magphys} is an integrated measure of its star formation. Therefore it represents the star formation activity of a galaxy
 averaged over a long time (specifically 0.1 Gyr for {\sc magphys}). But at $z\sim0$ it is linearly correlated with the instantaneous measure of SFR
 obtained from the H$\alpha$ emission line \citep[see fig.~1 of][]{mahajan18}.

 \section{Red vs Blue}
 \label{red}

 With a sample of red spiral galaxies defined, we now intend to explore the cause for the unusual optical colours for this subset of spiral galaxies in our sample. 
 
 \subsection{Viewing angle of red and blue spirals}
 
 Inclination of galaxies is known to correlate with optical colours \citep[e.g.][]{holmberg58,masters10a}. Using a sample of $\sim 24,000$ galaxies 
 from the SDSS, \citet{masters10a} 
 showed that not only is the effect of dust reddening significant on inclined spirals, but bulge-dominated early-type spirals are intrinsically red. 
 In order to test the impact of inclination in our sample of spirals, we first analysed the distribution of the inclination angle 
 of the red and blue spiral galaxies.
 If a spiral disk is represented by an oblate spheroid, the inclination $i$ of the plane of the galaxy to the line of sight is obtained by the relation
 \begin{equation}
 cos^2 i = \frac{(b/a)^2-q^2}{1-q^2},
 \end{equation}
 where $b/a$ is the observed ratio of minor to major diameters, and $q$ is the intrinsic axial ratio of the spheroid \citep{holmberg58}. In this work we use the 
 $q$ values listed in table~1 of \citet{masters10a} for different morphological types of galaxies as following:
 \begin{itemize}
 \item S0-Sa, SB0-SBa :  0.23
 \item Sab-Scd, SBab-SBcd :  0.20
 \item Sd, Unclassified :  0.103
 \end{itemize}
 Since typically $q\lesssim0.2$, we note that the distributions does not change qualitatively even if we fix $q$ to this value.  
 
  \begin{figure}
 \centering{
 {\rotatebox{270}{\epsfig{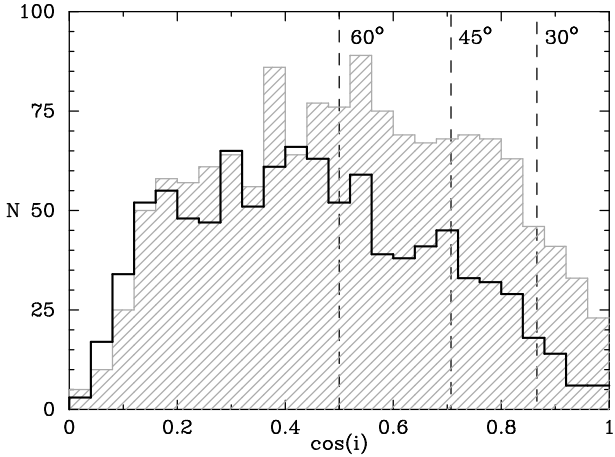}}}}
 \caption{The cos(i) distribution for red {\it (solid line)} and blue {\it (hatched)} spiral galaxies in our sample show that statistically, the red spiral galaxies are more
 inclined compared to blue spiral galaxies. }
 \label{cos}
 \end{figure}

 In this work we use the $a/b$ ratio in the $r$-band to determine the inclination as defined above. If our sample is unbiased, it should have a flat 
 distribution of $cos(i)$ signifying randomly oriented galaxies. Fig.~\ref{cos} however shows that the truth is far from this assumption. Firstly, we notice that
 the distribution of red and blue spiral galaxies with $i>60^{\circ}$ is statistically similar, but at inclination angles below $60^{\circ}$, the median 
 $i$ of red galaxies is likely to be $3^{\circ}$ more than their blue counterparts. We confirm the difference between the inclination of red and blue
 spirals using the Kolmogorov-Smirnov (KS) statistic, which tests for the null hypothesis that the two distributions are drawn from the same parent sample. 
 In this case the KS statistical probability is $1.25e-10$, thus rejecting the hypothesis that the inclination of red and blue galaxies is similar.
 Furthermore, while 51\% of the blue spiral galaxies have $cos (i) < 60^{\circ}$, only 37\% of the red spirals follow suit. These observations suggest that optical 
 colours of at least some of the red spirals may be due to their inclination relative to the line of sight.
 
 Figure~\ref{cos} also shows that our sample has a deficit of galaxies with $i < 35^{\circ}$ and $i \gtrsim 84^{\circ}$. This is in broad agreement with the  
 observations of \citet{masters10b} who found a similar deficit of galaxies with $i < 25^{\circ}$ and $i > 84^{\circ}$. Moreover, we find the distribution of $cos (i)$ 
 for the red and blue spiral galaxies to be relatively flat in the range $56^{\circ} < i < 84^{\circ}$ and $35^{\circ} < i < 70^{\circ}$, respectively. 
 
 In order to test if there is any correlation between the \gr~colour and $cos (i)$ we use the 
 Spearman's rank correlation statistic, which tests for the strength and direction of the monotonic relation between two variables. We find the Spearman's rank correlation and the corresponding
 probability ($p$) that the rank deviates from zero to be $-0.038~(p=0.15)$ and $-0.063~(p=0.05)$ for the blue and red spiral galaxies, respectively. This result confirms that there is no 
 significant correlation between the \gr~colour and $cos (i)$ for the blue spiral galaxies, and only a marginal correlation for the red spiral galaxies. The latter is expected due to optical reddening of
 the edge-on spiral galaxies as discussed in detail by \citet{masters10a}.
  
  \begin{figure}
 \centering{
 {\rotatebox{270}{\epsfig{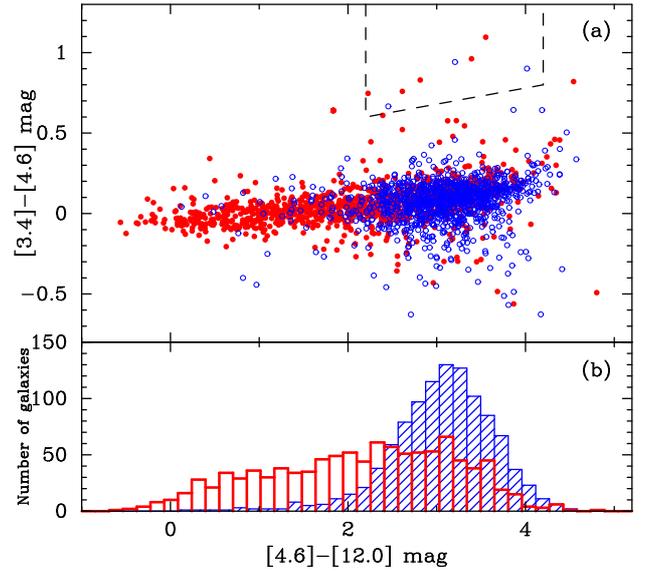}}}}
 \caption{This figure shows the (a) distribution of red {\it (solid red points)} and blue {\it (open blue circles)} spiral galaxies in the \w~colour-colour plane, and (b) the distribution of 
 the [4.6]-[12] colour for the red {\it (solid line)} and blue {\it (hatched)} spiral galaxies, respectively. The dashed region represents the 
 colour space where AGN are 
 expected to lie (see text). The \w~colours of both the colour-selected samples follow the locus expected for nearby spiral galaxies \citep{jarrett11}. Optically red spiral galaxies 
 have bluer \w~colours indicating lower SFR as compared to the blue spiral galaxies, although the red optical colour of some of the red spirals in our sample can be explained by 
 dust obscuration. On the other hand, optically blue spiral galaxies occupy the high SFR end of the \w~colour space, with a few of them having \w~colours expected for 
 starburst galaxies. }
 \label{wise}
 \end{figure}
  
  \subsection{Dust content}
  
  In this work we make use of the \w~photometry for all our galaxies to test if the red spirals are optically reddened due to dust. \w~infrared data are available for
  around 95\% of the blue and 98\% of the red spiral galaxies for our sample. The \w~colours are 
  subtly different from those obtained from other infrared observatories such as {\it Spitzer} because of the 12.0 \m m (W3) band. The W3 band is sensitive 
  to the poly-cyclic aromatic hydrocarbon (PAH) emission at 11.3 \m m from the nearby galaxies, as well as warm continuum emission 
  from active galactic nuclei (AGN) at all redshifts \citep{jarrett11}. While stars and elliptical galaxies have \w~colours $\sim 0$, spiral galaxies are red in [4.6]-[12], and
  ultra-luminous infrared galaxies (ULIRGs) tend to be red in the [3.4]-[4.6] colour as well. 
  
  In Fig.~\ref{wise} we show the colour-colour distribution of the red and 
  blue spiral galaxies, respectively. This figure can be directly compared to fig.~26 of \citet{jarrett11} or fig.~12 of \citet{wright10}\footnote{In order to 
  make this comparison, Fig.~\ref{wise} shows colours in Vega magnitude unlike all other colours and magnitudes discussed in this paper which are expressed as AB
  magnitude.}. It is interesting to note that despite their optical colour, the red spirals occupy the region of the diagram expected for nearby spiral galaxies. 
  On the other hand, blue galaxies are found in a cloud at [4.6]-[12]$ \sim 4$ mag expected for spiral galaxies with high SFR, and a small fraction occupying colour space 
  ([4.6]-[12]$ > 4$ mag and  [3.4]-[4.6]$ \sim 0.2$ mag) expected for starburst galaxies. Since redder [4.6]-[12] colour implies higher dust content, Fig.~\ref{wise} suggests that 
  the blue spiral galaxies are more dusty than their red counterparts. Specifically, while 95\% of the blue spiral galaxies have [4.6]-[12]$ > 2$ mag, about 58\% of the red spiral
  galaxies follow suit. In agreement with other such studies \citep{cluver14}, we find that irrespective of the optical colour, \w~colours of very few galaxies in our sample 
  fall in the range expected for AGN. 

 Therefore in the light of the facts that (i) the blue and red spiral galaxies with redder \w~colours ([3.4]-[4.6]$ \gtrsim 0$ and [4.6]-[12]$ > 2$ mag), have overlapping
 infrared properties and, (ii) all spiral galaxies irrespective of their optical colours show no tendency to be hosting an AGN, neither dust nor AGN activity 
 can fully explain the cause for the optical colours of the red spiral galaxies.

 \subsection{Environment of red spirals}
 
 \begin{figure}
 \centering{
 {\rotatebox{270}{\epsfig{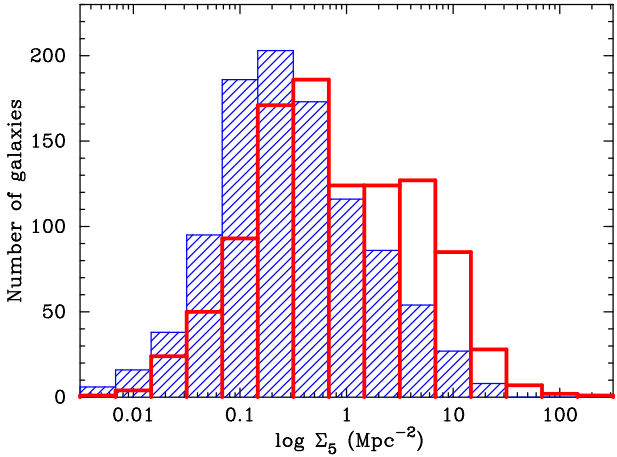}}}}
 \caption{The distributions of the projected galaxy density $\Sigma_5$ for the red ({\it open histogram}) and blue ({\it shaded histogram}) spirals show that the red spiral 
 galaxies are more likely to be found in high density regions relative to their blue counterparts.}
 \label{env}
 \end{figure}
  
 In this paper we quantify the environment of galaxies by the nearest neighbour surface density parameter, $\Sigma_5$. For a galaxy {\it G}, $\Sigma_5$ is defined as 
 the projected density of galaxies within a circle centred at {\it G} having radius equal to the distance to the fifth nearest neighbour to {\it G}. We use the $\Sigma_5$ 
 estimates from the EnvironmentMeasures DMU v05 \citep{brough13}. \citet{brough13} estimated $\Sigma_5$ using projected comoving distance to the fifth nearest neighbour
 within $\pm1000$ km s$^{-1}$. The density-defining population is also required to have the absolute SDSS petrosian magnitudes $M_r$ less than the limiting magnitude
 of $M_{r, limit}=-20.0$ mag. Galaxies where the nearest survey edge is closer than the fifth nearest neighbour are flagged, and have only the upper limits assigned to them.
 
 $\Sigma_5$ values were obtained for 1399/1463 (96\%) blue and 1027/1034 (99\%) red spiral galaxies, respectively and their distributions are shown in Fig.~\ref{env}. 
 As a comparison, only 5\% of the blue spiral galaxies have $\Sigma_5 > 5$ Mpc$^{-2}$ relative to 16\% of the red spirals, clearly indicating that spirals found in high 
 density environments are more likely to be red. We also confirm that the redshift distribution of the red and blue spiral galaxies are statistically similar, and hence do not
 effect the distribution of $\Sigma_5$. 

  \begin{figure}
 \centering{
 {\rotatebox{270}{\epsfig{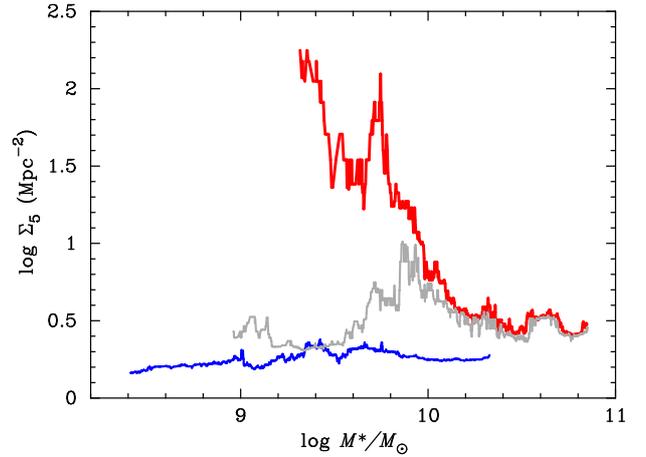}}}}
 \caption{This figure shows the median distribution of the projected galaxy density $\Sigma_5$ for the red ({\it thick red line}) and blue ({\it thin blue line}) spiral galaxies as a function of their 
 stellar mass. The {\it grey} distribution shows the same for all the spiral galaxies taken together. It is evident that at fixed stellar mass, red spiral galaxies on average prefer higher density 
 environments. Furthermore, counter-intuitively the stellar mass of red spiral galaxies is anti-correlated with their environmental density. }
 \label{mass-env}
 \end{figure}
 
 In Fig.~\ref{mass-env} we show the median of log $\Sigma_5$ as a function of stellar mass in running bins for the two colour-selected populations of spiral galaxies. Despite the 
 statistically distinct distributions of stellar mass for the red and blue spiral galaxies (see Fig.~\ref{magphys} and Table~\ref{ks:red}), it is evident that at fixed stellar mass, especially at 
 log $M^* \lesssim 10^{10.5} M_\odot$, the red spiral galaxies always reside in denser environments relative to their blue counterparts. 
 It is also interesting to note that stellar mass of red spiral galaxies is anti-correlated with their environmental density, such that the more massive red spirals reside in lower density 
 environments relative to their less massive counterparts. The blue spiral galaxies on the other hand are always found in low-density environments.
 
 Together, Figs.~\ref{env} and \ref{mass-env} support the hypothesis that spiral galaxies are affected by some environmental mechanism(s) in dense
 environments, leading to stripping of gas or a burst of star formation, both of which will eventually quench formation of stars due to loss of gas. 
    
 \subsection{Physical properties of red and blue spiral galaxies}
 
 Physical properties like SFR, sSFR and \smassa of galaxies are known to correlate with their colour. In this section we explore these physical 
 observables of red and blue spiral galaxies and test for their correlation with optical colour.
   
 \begin{figure*}
 \centering{
 {\rotatebox{0}{\epsfig{file=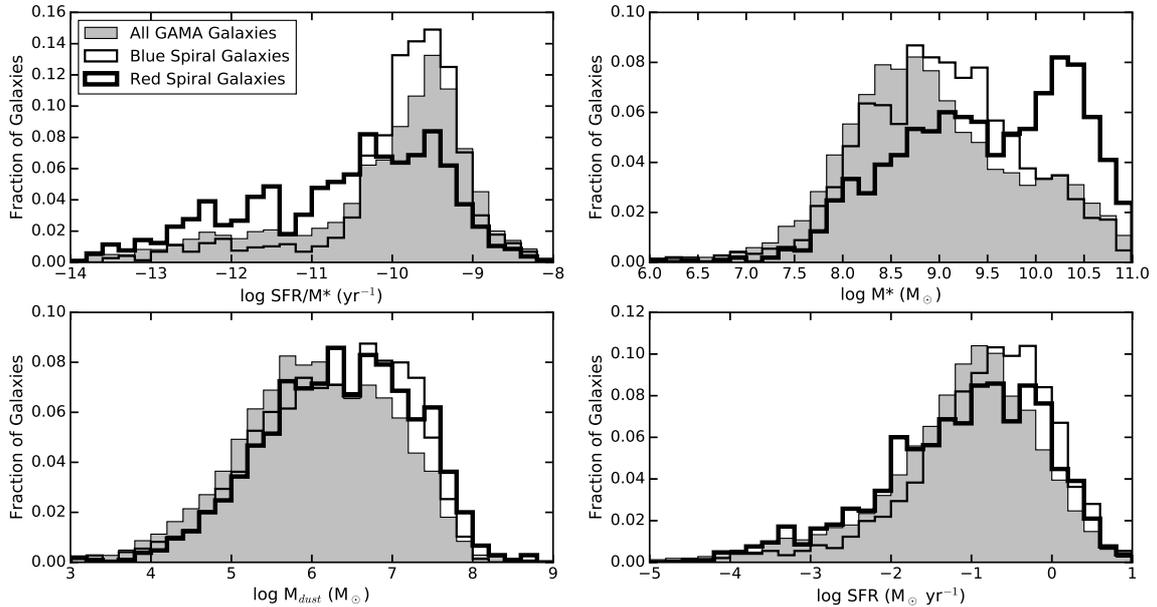,width=18cm}}}}
 \caption{A comparison between {\it (clockwise from top left)} (a) sSFR, (b) \smassa, (c) SFR and (d) \dmass~for the red and blue spiral galaxies in our sample, and all the
  galaxies ($0.002<z<0.06$) in the \g catalogue. The red spiral galaxies are found to have lower sSFR and higher \smassa~relative to blue spiral galaxies. Therefore, even though the
  \dmass~and SFR distributions for the two colour-selected samples seem similar, at fixed stellar mass red spiral galaxies will have lower star formation rate and dust masse than 
  comparable blue spiral galaxies (also see Fig.~\ref{dts-dist}). }
 \label{magphys}
 \end{figure*}

 In Fig.~\ref{magphys} we show a comparison between various physical properties of the red and blue spiral galaxies, and Table~\ref{ks:red} shows the
 KS statistical probability for the two distributions to be drawn from the same parent sample. The blue and red spirals have statistically significantly different 
 \smassa, SFR and sSFR, viz. the red spirals are forming less stars and are more massive relative to their blue counterparts. These results are in agreement with the previously 
 published literature based on morphologically-selected samples of spirals \citep[e.g.][]{masters10b}. 
 
 Fig.~\ref{magphys} also shows that the distributions of  dust mass among the red and blue spirals are similar. These observations can be explained using the SFR--$M_{dust}$ relation from the 
 literature. \citet{dacunha10} have quantified the SFR-\dmass~relation for low redshift SDSS galaxies as $M_{dust} \sim SFR^{1.11\pm0.01}$. 
 In a study of the \smassa-\dmass~relation using a complimentary sample of galaxies also from the SDSS, \citet{hjorth14} discussed various probable causes of such an observation. 
 Amongst others, \citet{hjorth14} proposed that in the initial stages of a starburst, SFR and dust mass increases together, giving rise to the SFR-$M_{dust}$ correlation. Thereafter, if a galaxy is
 quenched through removal of cold gas and dust from the galaxy, both SFR and \dmass~will decline together leading to a transition parallel to the SFR-\dmass~relation. Therefore, it is
 plausible for a massive, red spiral galaxy to have the same amount of dust as a less evolved blue spiral galaxy. However, if the SFR in a galaxy declines, but dust is retained 
 \citep[e.g.][]{martig09,genzel14}, a galaxy will transit horizontally in the SFR-\dmass~plane. On the other hand, mergers will cause a vertical upward transition, conceivably leading to a 
 secondary burst of star formation \citep{hjorth14}. Since many of our red spiral galaxies are massive (Fig.~\ref{magphys}), and, at a given stellar mass redder compared to their blue counterparts, 
 our data supports the former scenario. But we do not find disturbed morphology in a statistically significant fraction of red spiral galaxies to support the latter hypothesis. 
    
 \begin{table}
 \caption{The KS test statistical probability for the likelihood that the red and blue spiral galaxies are derived from the same parent sample.}
 \begin{center}
 \begin{tabular}{ c|c }     
 \hline
      Distribution  & KS statistical probability  \\ \hline
        SFR &  1.18 e-09  \\ 
       sSFR   & 1.96 e-64 \\ 
    \smass & 1.10 e-35   \\ 
    \dmass & 1.72 e-01  \\ 
  \hline
 \end{tabular}
 \end{center}
 \label{ks:red}  
 \end{table} 
 
 
  \begin{figure}
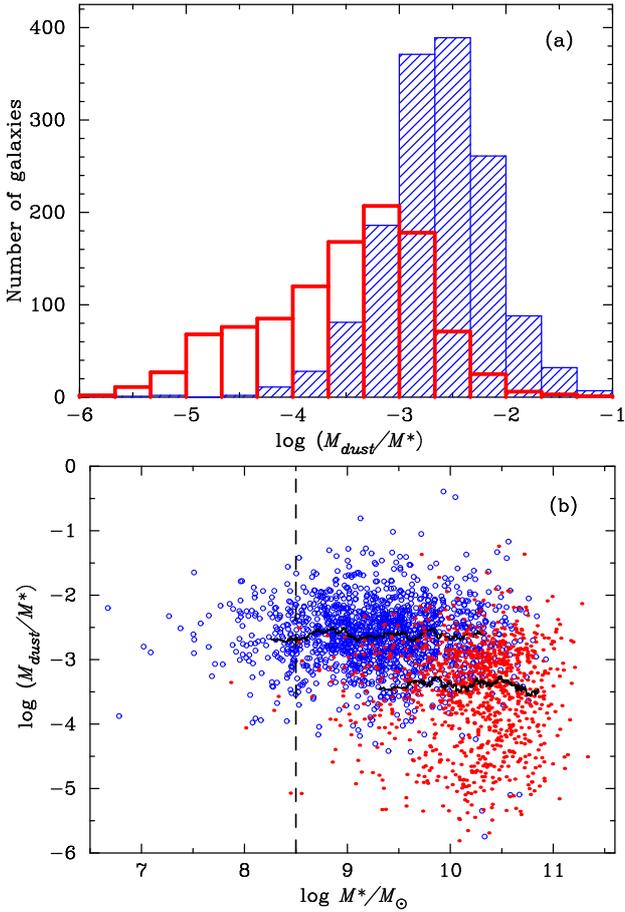

  \centering{
 {\rotatebox{270}{\epsfig{file=dts3.ps,width=6cm}}}}
  \centering{
 {\rotatebox{270}{\epsfig{file=dts-smass5.ps,width=6cm}}}}
   \caption{(a) The distribution of dust-to-stellar-mass ratio for the blue {\it (hatched)} and red {\it (solid)} spiral galaxies, respectively. The specific production of dust is lower in the red 
   spiral galaxies as compared to the blue spiral galaxies. (b)  The DTS ratio is shown as a function of \smassa~for the two colour-selected samples of spiral galaxies. Symbols are same as 
   in Fig.~\ref{wise}. The {\it horizontal solid horizontal lines} represent the median trends in the two colour-selected samples, while the {\it vertical dashed line} represents the limiting stellar 
   mass at the maximum redshift of our sample.} 
 \label{dts-dist}
 \end{figure}
 
 In Fig.~\ref{dts-dist} (a) we show the distribution of the dust-to-stellar mass (DTS) ratio for the red and blue spirals in our sample. The specific production of dust in the red spirals is notably
 lower than their blue counterparts as indicated by a shift in the mean of the DTS distribution. 
 For different samples of spiral galaxies in literature, the DTS ratio is found to be anti-correlated with \smass~in different environments \citep{cortese12a, calura17}. This observation is 
 considered as an indication of a scenario where the balance between dust production and destruction is dependent on the stellar mass of a galaxy. But as shown in Fig.~\ref{dts-dist} (b), we do 
 not observe such a trend. On the contrary, we observe that within each colour-selected population of spiral galaxies, the DTS ratio remains constant with stellar mass of galaxies
 such that the median DTS ratio for the red spirals is lower by $\sim 1$ dex than their blue counterparts. This result is in broad agreement with the 
 results of \citet{rowlands12}, who found that the passive spiral galaxies detected in the Herschel-Astrophysical Terahertz Large Area Survey (H-ATLAS) data have lower DTS ratio, higher 
 \smassa~and older stellar population ages as compared to normal spiral galaxies.


 \section{Discussion}
 \label{discuss}
 
 The aim of this paper is to examine red spiral galaxies found in the \g sample. We make use of several galaxy properties such as 
 \smassa, SFR and dust mass to investigate the cause for the optical colours of red spiral galaxies.
 In the following we discuss the existing literature in the context of our findings, and list some of the major caveats in our analysis.
 
  \subsection{Star formation in red spiral galaxies}
 
  \begin{figure*}
  \centering{
 {\rotatebox{270}{\epsfig{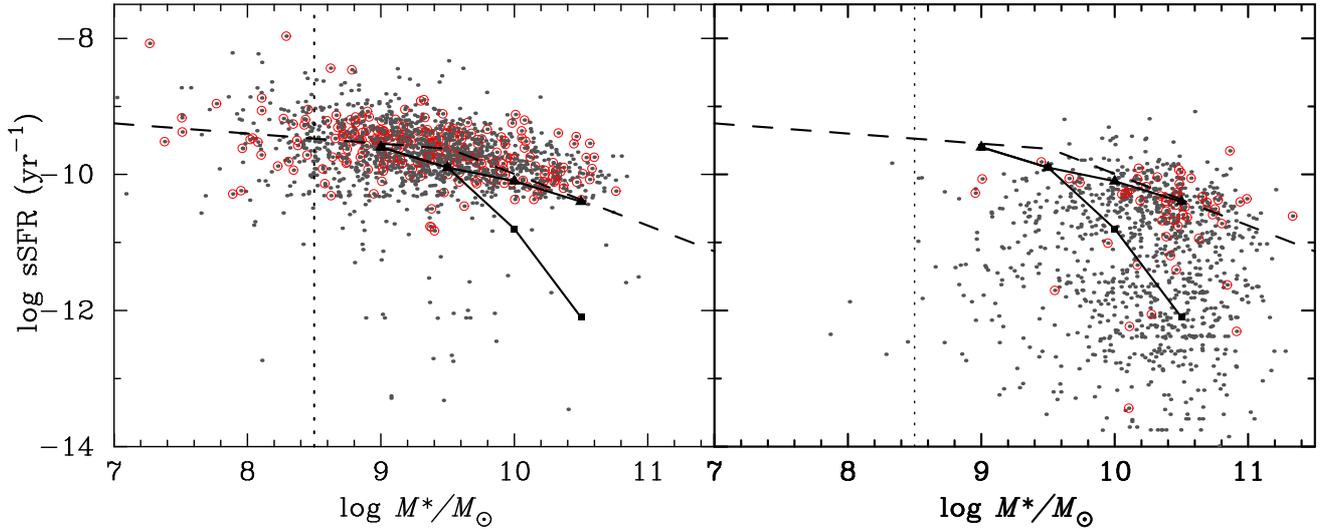}}}}
  \caption{The sSFR as a function of \smassa~for the {\it (left)} blue and {\it (right)} red spiral galaxies, respectively. The encircled points represent galaxies detected in HI and the {\it vertical dotted line}
  represents the mass completeness limit at the highest redshift of our sample. The {\it solid lines and squares} show the relation obtained for star-forming galaxies ($0.05<z<0.11$) by \citet{bauer13}, 
  while the {\it triangles and the corresponding lines} represent the same for their complete GAMA sample. Another relation obtained by \citet{huang12} for an HI-selected sample of 
  local galaxies is shown by {\it dashed lines}.}
 \label{ssf-lgm}
 \end{figure*}

 In the optical waveband, the sSFR-\smassa~relation for a general population of galaxies shows a sharp decline, such that massive galaxies have lower sSFR \citep[e.g. fig.~2 of][]{bauer13}. 
 On similar footing, a sample selected from the submillimeter surveys shows that all galaxies lie on a single, curved sequence of galaxies without there being a need for a separate 
 main galaxy sequence for star-forming galaxies \citep{eales18a,eales18b}.
 In Fig.~\ref{ssf-lgm} we re-confirm this well established trend by showing the sSFR of blue and red spirals as a function of their \smassa. We also show the HI-detected\footnote{Only 
 11\% of blue and 5\% of red spiral galaxies are detected in HI by the ALFALFA survey. For further details see Appendix~\ref{s:h1}.} galaxies in both the 
 colour-selected sub-samples. There are two subtle features in this figure. Firstly, the colour-selected samples of spirals are distributed very differently in the \smassa--sSFR plane. 
 The blue galaxies span the entire \smassa~range covered by this sample, and exhibit a linear anti-correlation with the sSFR. On the other hand, the red spiral galaxies 
 are distributed randomly at the high mass end and almost always have sSFR $\lesssim 10^{-10}$ yr$^{-1}$, i.e. the sSFR of red spiral galaxies is typically below that of their blue counterparts, 
 at a given \smassa. 
  
 The distribution for our blue spiral galaxies in the sSFR-\smassa~plane agrees very well with the relation derived by \citet{bauer13} for star-forming 
 (EW(H$\alpha$)$> 3$\AA\ and ${\rm F}_{H\alpha} > 2.5 \times 10^{-16}$ erg s$^{-1}$ cm$^{-2}$ \AA$^{-1}$) galaxies in their GAMA sample ($0.05<z<0.11$), while the red spiral galaxies 
 seem to follow the relation 
 they derived for `all' galaxies. Furthermore, \citet{gavazzi15} have compared the  sSFR-\smassa~relation for their sample of HI-detected local galaxies with the distributions derived 
 by \citet{bauer13} and \citet{huang12}, also shown in Fig.~\ref{ssf-lgm}, and found them to agree well within uncertainties, just like the distributions shown here. 

\begin{figure}
  \centering{
  {\rotatebox{270}{\epsfig{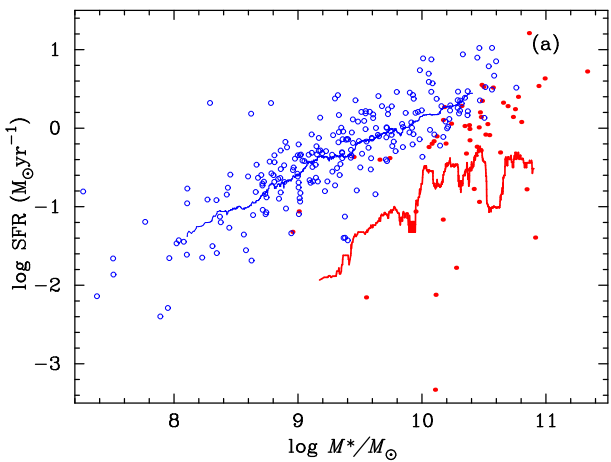}}}}
 \vspace{0.2in}
  \caption{ (a) In this figure we show the median SFR of galaxies as a function of their stellar mass for the HI-detected galaxies. Symbols are same as in Fig.~\ref{wise}. The median SFR of red spiral 
  galaxies is found to be less than their blue counterparts by $\sim 1$ dex at all stellar masses.}
 \label{sfe}
 \end{figure}

 Even though only a small fraction of our spiral galaxies are detected in HI, it is insightful to explore the distribution of these galaxies in the SFR-\smassa plane, which is shown in Fig.~\ref{sfe}. 
 As observed by several other authors \citep{huang12,parkash18,zhou18}, we also find that the SFR is correlated with the stellar mass of galaxies, such that the SFR increases with increasing 
 \smassa. The median SFR of the red spiral galaxies is lower by $\sim 1$ dex relative to their blue counterparts over almost two orders of magnitude in stellar mass. This result, together with 
 Fig.~\ref{mass-env} evidently shows the impact of mass quenching even among the massive gas-rich spiral galaxies. 
 Fig.~\ref{sfe} also suggests that a significant fraction of the observed scatter in the star formation main sequence \citep{bauer13,grootes13,speagle14,parkash18}  
 may result from the treatment of passively-evolving or red disk galaxies in a sample. This observation supports the result of \citet{parkash18} who showed that the scatter in the \smassa-SFR 
 relation is anti-correlated with the T-type of spiral galaxies in their HI-selected sample. 

 The results presented in this subsection, together with Fig.~\ref{mass-env} suggest that the red spirals have gained their optical colour by losing their gas via some mechanism, which in turn 
 lead to a reduction in the rate at which they were forming stars.

 \begin{table*}
 \caption{The sky coordinates, $z$, {\it (g-r)} colour, ALFALFA ID and HI mass of spiral galaxies used in this work. (A complete version of this table is available online). }
 \begin{center}
 \begin{tabular}{ ccccccc }     
 \hline
  GAMA ID & Right Ascension & Declination & Redshift & {\it (g-r)} & ALFALFA ID & $M_{HI}$ \\
                  & (J2000)              & (J2000)      &                  & mag   &                 & ($M_\odot$) \\  \hline
   7623 & 178.509 &   0.785 & 0.053 &  0.445 &   -   &  -  \\ 
   7802 & 179.443 &   0.804 & 0.048 &  0.333 &   -   &  -  \\ 
   8217 & 181.052 &   0.803 & 0.012 &  0.396 &  223256 &  0.35E+09 \\ 
   8353 & 182.017 &   0.698 & 0.021 &  0.415 &  222071 &  0.19E+10 \\ 
   8706 & 183.689 &   0.744 & 0.022 &  0.582 &    7250 &  0.33E+10 \\ 
   9061 & 184.824 &   0.824 & 0.049 &  0.367 &   -   &  - \\ 
   9064 & 184.817 &   0.648 & 0.023 &  0.535 &   -   &  - \\ 
   9112 & 184.951 &   0.733 & 0.043 &  0.408 &   -   &  - \\ 
   9115 & 184.965 &   0.810 & 0.035 &  0.480 &   -   &  - \\ 
   9163 & 185.141 &   0.788 & 0.009 &  0.416 &    7396 &  0.11E+10 \\ 
 \hline
 \end{tabular}
 \end{center}
 \label{galaxies}  
 \end{table*}

 \subsection{Red spirals in the literature}
 
 Passively-evolving spiral galaxies have traditionally been studied as a transitional galaxy population, particularly in dense environments \citep[e.g.][]{vanden76}. 
 But lately different selection criteria have resulted in mutually exclusive samples of optically-red spiral galaxies with distinguishable star formation properties. For instance, 
 a morphologically-selected sample of disky red spiral galaxies in the optical waveband \citep{masters10b}, has almost no overlap with a sample of passive spiral galaxies 
 selected in the $K-$band from the 2-micron all sky survey (2MASS) \citep{fraser16,fraser18}. 
 Due to the different sample selection criteria used by these authors, the sample of \citet{masters10b} includes dusty galaxies, but excludes the ones with high bulge-to-disk (BD) ratio, 
 while Fraser et al.'s selection criteria based on \w~colours eliminated dusty galaxies from their sample, but the T-type selection criteria included spiral galaxies with high BD 
 ratio.
 
 It is therefore essential to understand that red disc galaxies are a collection of several individual populations which are a product of different formation mechanism(s) or effects. For instance,
 dust reddening is important for edge-on disc galaxies \citep{masters10b}, while low SFR will lead to red optical colours irrespective of orientation \citep[e.g.][]{goto03,mahajan09}. On the other 
 hand, mass quenching becomes important for spiral galaxies with $M^*\gtrsim 3\times 10^{10} M_\odot$, and quenching due to environment will effect low mass spiral galaxies in dense 
 environments (Fig.~\ref{mass-env}), especially if they are satellites of larger galaxies \citep{haines06}.     
 Hence, such differences in the selection criteria must be taken into account while making a comparison between different studies.
 
 While in dense environments `dusty star-forming' and `passively-evolving' spiral galaxies appear to be the same phenomenon \citep{wolf09}, in a generic sample of galaxies 
 the evolution is governed by the availability of gas reservoir to form stars. 
 Using data from the SDSS and {\it Galex}, together with the morphological information from the Galaxy Zoo, \citet{schawinski14} 
 suggested that passive spirals are a result of slow quenching of star formation \citep[also see][]{cortese09,wolf09,koyama11,haines15}. On the other hand, morphological transformation 
 occurs when the star formation in a 
 galaxy is truncated on a very short time-scale, perhaps as a result of a merger event. This is also in broad agreement with the results of \citet{bremer18} who studied the morphological 
 transformation of the transitional green valley galaxies ($z<0.2$; $10.25 < {\rm log}~M^*/M_\odot < 10.75$) from the GAMA survey. \citet{bremer18} find that the transition time of 
 galaxies through the green valley is $\sim 1-2$ Gyr, and is independent of environment. 
 Their result is in agreement with the trend seen in Fig.~\ref{mass-env}, where the red spiral galaxies with log $M^*/M_\odot$ in the range $10.25-10.75$ in our sample do not show any significant 
 change in the environmental density. 

 In their work exploring the colours and morphology of transitional galaxies, \citet{bremer18} also find that 
 the green valley galaxies have significant bulge and disk components, but the transition from blue to red optical colour is driven by the colour change of the disk. 
 Their results led these authors to suggest that star formation is quenched in the disk as the gas content is used up, or becomes less available in the period following growth of the bulge. 
 Using data from the COSMOS survey, \citet{bundy10} also find that as much as 60\% of galaxies moving to the red sequence undergo a passive-disk phase.

 The BD ratio for a small sub-sample of 
 our galaxies (5\% red and 15\% blue spiral galaxies) are available from the BDDecomp DMU \citep{robotham17,robotham18}. Therefore, although we are unable to provide a complete
 analysis comparable to \citet{bremer18} for our sample, we do find indications suggesting that the BD ratio for the red spiral galaxies is statistically larger than their 
 blue counterparts, with the median (standard deviation) BD ratio being 0.82 (5.05) for the red and 0.46 (8.13) for the blue spiral galaxies, respectively. But the existence of red spiral galaxies
 with small BD ratio in our colour-selected sample suggests that 
 in agreement with the literature \citep{bundy10}, fading of the blue disk alone is insufficient at explaining the origin of passive disk galaxies.
       
 In a recent work \citet{evans18} studied a population of red star-forming galaxies ($z>0.05$, ${\rm log}~M^*/M_\odot > 9.5$) in the local universe using the data from the SDSS 
 (data release 7). In line with the literature \citep[e.g.][]{mahajan09}, they found that $\sim 11\%$ of galaxies at all stellar masses are optically red, yet forming stars. Unlike the work 
 presented here however, \citet{evans18} find that the 
 proportion of their `red misfit' galaxies is independent of environment, where the latter is quantified by group-centric distance as well as the group halo mass. These authors, 
 along with several others \citep{masters10b,salim14,gu18} also find that emission-line `red disk' galaxies are more likely to host an optically-identifiable AGN relative to 
 their blue counterparts \citep[but see,][for an alternate view]{fraser16}. However, since we have only used \w~colours to identify AGN (Fig.~\ref{wise}) in this work, a direct comparison of our results 
 with \citet{evans18} is not feasible.
  
 In a nutshell, although red disk galaxies comprise of different sub-populations, literature seems to indicate that at least in the nearby Universe ($z\lesssim0.1$) optically-red disk galaxies
 are more dusty, less star-forming and more massive than their blue counterparts. Most of the discrepancies among different papers seem to be a result of different selection criteria
 used to select the respective samples. 
  
 \subsection{Causes of caveats in our analysis}
  
  \begin{itemize}
  \item 
  {\it Inclination of galaxies}: In the analysis presented here we have included disk galaxies with a range of inclination (Fig.~\ref{cos}) because we believe this is essential for good statistics
  in a study like the one presented here. 
  \item 
 {\it Dust in spirals}: While \citet{masters10b} excluded dusty galaxies in their analysis, other authors \citep[e.g.][]{wolf09,koyama11} have chosen differently. In this work
 we have included all spiral galaxies irrespective of their dust content, in order to avoid any biases in our sample. Fig.~\ref{wise} and 
 \ref{dts-dist} validate our methodology of incorporating all galaxies irrespective of their dust content, because otherwise many of the differences observed in the dust properties of the 
 colour-selected samples of spirals would not have been observed.      
  \item
  {\it Poor resolution of optical images}: The morphological classification used in this work comes from the \g~DMU which employs the SDSS imaging data. It is therefore
  noteworthy that despite careful classification and multiple attempts it is likely that at least some spiral galaxies, especially the ones having low SFR may have been 
  misclassified as an elliptical in the shallow $\sim 55$ seconds long exposure images \citep{bonne15}. We therefore suggest the reader to use the fractions and numbers quoted in 
  this work cautiously. 
  \item
 {\it Limited sensitivity of HI data}: As discussed in Appendix~\ref{s:h1}, since HI surveys are designed to detect galaxies rich in HI, they are unfair representatives of gas-poor galaxies. The 
 correlation between atomic gas mass and stellar mass further implies that a high limiting \hmass~will in turn reduce the effective sensitivity of the HI-detected sample relative to an optical 
 selected one. The HI-analysis presented in this work is therefore only limited to massive, gas-rich spiral galaxies, which may not be a true representations of the entire population of spiral galaxies.      
  \end{itemize} 
 
\section{Summary}
\label{summary}
 
 In this paper we examine various properties of morphologically-selected spiral galaxies to test for the origin of red colours of some of them. 
 To accomplish this we used the optical data and its derivatives compiled for low-redshift ($0.002<z<0.06$) galaxies in the GAMA survey, and for a small sub-sample of those, HI data
 obtained from the ALFALFA survey. Specifically, we probed the viewing angle and physical properties: SFR, sSFR, \smassa~and 
 \dmass~of the red and blue spiral galaxies. The main results of this work are:
 \begin{itemize} 
 \item The distribution of inclination angle of the red and blue spirals is different, yet do not correlate with optical colour or other physical properties suggesting that 
 the red colours of spirals in our sample do not originate because of their viewing angle alone.
 \item The \w~colours of all spiral galaxies in our sample comply with the range of colours expected for nearby star-forming galaxies, although optically blue spiral galaxies have
 redder \w~colours indicating high SFR. A small fraction of optically-red spiral galaxies also exhibit blue \w~colours suggesting that dust obscuration may have caused red colours 
 in some of them.
 \item The SFR, sSFR and \smassa distributions of red and blue spirals are statistically different, yet their \dmass~distributions are similar. 
 \item Spiral galaxies in dense environments are more likely to be optically red. Furthermore, at fixed \smassa~red spiral galaxies preferentially reside in high density environments 
 relative to their blue counterparts. 
 \item The dust-to-stellar mass ratio for spiral galaxies is independent of \smassa~within each colour-selected population. But the DTS ratio for the red spiral galaxies is lower by 
 $\sim 1$ dex relative to their blue counterparts at all stellar masses. 
\end{itemize}

 To conclude, our results suggest that optical colour of red spiral galaxies are a resultant of several different effects and phenomenon. While the edge-on disc galaxies may 
 appear red due to inclination effect, a small but appreciable fraction of spiral galaxies have acquired red colours due to dust. The remaining population of red spirals seems
 to be a product of environmental effects which lead to loss of gas and dust, and eventually low SFR. 
 Observations supporting the impact of such environmental mechanisms on the transformation of galaxies in intermediate-density environments of galaxy groups \citep{rasmussen12} 
 and large-scale filaments \citep{mahajan12, mahajan18} have been discussed in the literature. In particular, the results of \citet{rasmussen12} favour a quenching timescale of 
 $\gtrsim 2$ Gyr, which is in broad agreement with the suggested scenario for the creation of passively-evolving spirals by other studies \citep{wolf09, koyama11, schawinski14}, and 
 the transition time of passive disc galaxies through the green valley \citep{cortese09,bremer18}.

\section{Acknowledgements}
 
 GAMA is a joint European-Australasian project based around a spectroscopic campaign using the Anglo-Australian
 Telescope. The GAMA input catalogue is based on data taken from the Sloan Digital Sky Survey and the UKIRT Infrared
 Deep Sky Survey. Complementary imaging of the GAMA regions is being obtained by a number of independent
 survey programmes including GALEX MIS, VST KiDS, VISTA VIKING, WISE, Herschel-ATLAS, GMRT and ASKAP
 providing UV to radio coverage. GAMA is funded by the STFC (UK), the ARC (Australia), the AAO, and the
 participating institutions. The GAMA website is http://www.gama-survey.org/ . This research made use of the 
 ``K-corrections calculator'' service available at http://kcor.sai.msu.ru/ .

 Mahajan is funded by the INSPIRE Faculty award (DST/INSPIRE/04/2015/002311), Department of Science and
 Technology (DST), Government of India. Gupta and Rana were supported by the INSPIRE scholarship for higher education for their BS-MS 
 dual degree at IISER Mohali. Rana was also partially supported by the DST/INSPIRE/04/2015/002311 during the course of this project.
 The authors are grateful to the anonymous reviewer for their criticism which greatly improved the clarity of this manuscript.
 
  \label{lastpage}


 \appendix
\appendixpage

 \section{Atomic gas mass in spirals}
 \label{s:h1}

 \begin{figure}
 \centering{
 {\epsfig{file=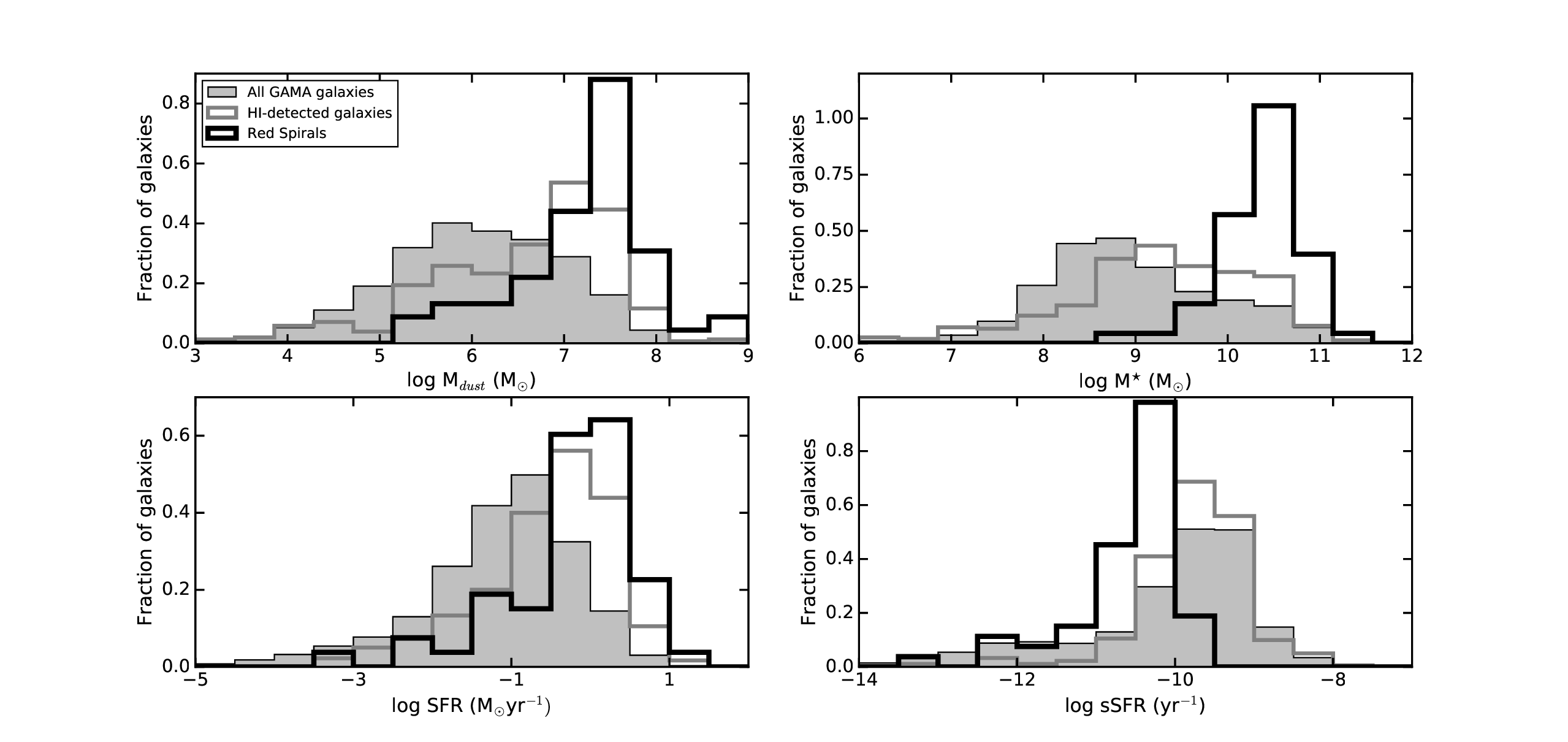,width=9cm}}}
 \caption{This figure shows the distributions of (a) sSFR, (b) \smassa, (c) \dmass and (d) SFR for all the HI-detected galaxies in our sample, all HI-detected spirals and the red spiral galaxies,
  respectively. The \smassa~distribution shows the effect of lower sensitivity of the HI survey relative to the optical data, i.e. most of the galaxies detected in HI are massive systems as 
  compared to a typical \g~galaxies in our redshift range. All but the pair of distributions of the HI-detected galaxies and HI-detected red spiral galaxies are found to be statistically 
  significantly different (Table~\ref{ks:h1}). }
 \label{magphys-h1}
 \end{figure}
 
 Star-forming galaxies may turn red in the absence of fuel for star formation. In order to test what fraction of spirals in our sample
 have reddened by exhausting their hydrogen gas content, we utilised the 21-cm continuum data from the Arecibo Legacy Fast Arecibo L-band Feed Array 
 survey \citep[ALFALFA;][]{giovanelli05}. ALFALFA is a blind extragalactic HI survey done using the Arecibo telescope to conduct a census of HI in the local universe. 
 Specifically, in this work we have used the $\alpha70$ catalogue \citep{haynes11}, which includes 70 per cent of the ALFALFA
 data\footnote{http://egg.astro.cornell.edu/alfalfa/index.php}. We matched our GAMA source list against the list of most probable optical counterparts chosen from the SDSS database 
 to check which galaxies are detected in HI. This exercise resulted in 361 sources matched within $3.5\arcsec$, of which 238 (66\%) are matched within $1\arcsec$. Of the 361 matched 
 galaxies, 287 (80\%) are spirals of which 53 (15\%) are red, which amounts to around 11\% of blue and only 5\% of red spiral galaxies in our sample. The ALFALFA ID and HI mass of all galaxies 
 in our sample, along with the relevant information from the GAMA survey data are presented in Table~\ref{galaxies}.
 
 To get further insight into the properties of the HI-detected spirals in Fig.~\ref{magphys-h1} we compare some physical 
 properties of all GAMA galaxies with the HI-detected galaxies and the red spirals among them. Table~\ref{ks:h1} shows the 
 KS test statistical probability in favour of the hypothesis. 
 Statistically, none but the SFR distributions for the blue and red HI-detected spirals are likely to have been drawn from the same parent sample. The stellar 
 mass distribution of the HI-detected galaxies shows the impact of the lower effective sensitivity of the HI survey relative to the effective sensitivity of the optical data. As a result of 
 this selection-bias in the HI survey, the HI-detected galaxies are mostly limited to massive, gas-rich systems, and not a fair representation of the gas-poor galaxies. We find that the 
 HI-detected red spirals are on average more massive, have higher dust mass and lower star formation activity relative to their blue counterparts. However, due to the inherent biases discussed 
 above, the authenticity of these trends remains to be tested.

 \begin{table}
 \caption{The KS test statistical probability for the likelihood that all the HI-detected galaxies, the HI-detected spirals, and the red spirals among them are derived 
 from the same parent sample.}
 \begin{center}
 \begin{tabular}{ c|c|c }     
 \hline
      Samples & Physical property  & KS statistical   \\ 
                     &                              &  probability \\ \hline
     All \g, Red spirals &  SFR &  3.66e-14 \\
     HI-detected spirals, Red spirals &    & {\bf 0.42} \\  \hline
     All \g, Red spirals & sSFR &  1.88e-13 \\ 
    HI-detected spirals, Red spirals &    & 1.25e-15 \\  \hline
    All \g, Red spirals & \smassa &  1.01e-24  \\  
    HI-detected spirals, Red spirals &    & 2.00e-13 \\  \hline
    All \g, Redspirals  & \dmass &   5.65e-19 \\ 
    HI-detected spirals, Red spirals &    & 9.94e-5 \\
  \hline
 \end{tabular}
 \end{center}
 \label{ks:h1}  
 \end{table}

\label{lastpage}

\end{document}